\documentclass{article}

\usepackage{arxiv}

\usepackage[utf8]{inputenc} 
\usepackage[T1]{fontenc}    
\usepackage{hyperref}       
\usepackage{url}            
\usepackage{booktabs}       
\usepackage{amsfonts}       
\usepackage{nicefrac}       
\usepackage{microtype}      
\usepackage{lipsum}
\usepackage{graphicx}
\graphicspath{ {./images/} }
\usepackage{multirow}
\usepackage{outlines}

\title{Is that a Duiker or Dik Dik Next to the Giraffe? Impacts of Uncertainty on Classification Efficiency in Citizen Science}

\author{
 Vinod Kumar Ahuja \\
  University of Nebraska at Omaha\\
  Omaha, NE 68182 \\
  \texttt{vahuja@unomaha.edu} \\
   \And
 Holly K. Rosser \\
  University of Nebraska at Omaha\\
  Omaha, NE 68182 \\
  \texttt{hrosser@unomaha.edu} \\
  \And
 Andrea Grover \\
  University of Nebraska at Omaha\\
  Omaha, NE 68182 \\
  \texttt{andreagrover@unomaha.edu} \\
}

\begin{document}
\maketitle
\begin{abstract}
Quality control is an ongoing concern in citizen science that is often managed by replication to consensus in online tasks such as image classification. 
Numerous factors can lead to disagreement, including image quality problems, interface specifics, and the complexity of the content itself.
We conducted trace ethnography with statistical and qualitative analyses of six Snapshot Safari projects to understand the content characteristics that can lead to uncertainty and low consensus.
This study contributes content categorization based on aggregate classifications to characterize image complexity, with analysis that confirms that the categories impact classification efficiency, and an inductively generated set of additional image quality issues that also impact volunteers' ability to confidently classify content.
The results suggest that different conceptualizations and measures of consensus may be needed for different types of content, and aggregate responses offer a way to identify content that needs different handling when complexity cannot be determined \textit{a priori.}
\end{abstract}


\section{Introduction}

\begin{figure*}[h]
\includegraphics[width=\textwidth, height=8cm]{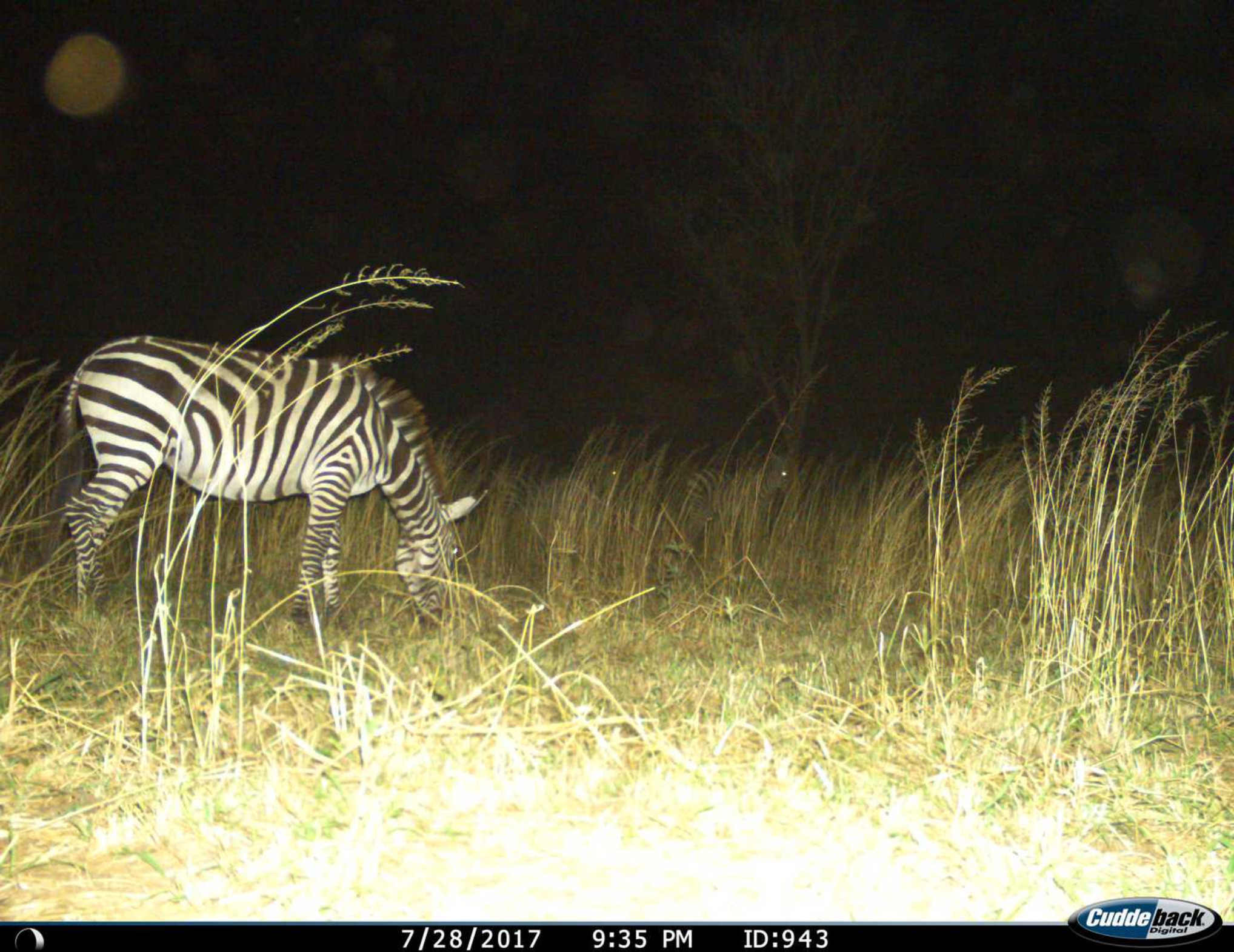}
\caption{Image content from Snapshot Grumeti. Here, the targets of the image are the zebras shown eating and resting.} \label{fig:Subject}
\end{figure*}

Researchers have longstanding concerns about data quality in citizen science due to the open nature of participation, and most projects employ multiple strategies for ensuring rigorous results \cite{Wiggins2015}.
In online image analysis projects, consensus on classifications by multiple human workers or volunteers is the most common strategy for ensuring data quality.
Often the image content for classification includes plants, animals, or visual representations of data, e.g., in astronomy. 
In the research presented here, the subjects of the images are animals in their natural habitats in African wildlife preserves (see Figure \ref{fig:Subject} for an example from Snapshot Grumeti).

Consensus on image classification is a complex data quality issue for citizen science projects in which volunteers classify the content of images through an online platform, such as the Zooniverse \cite{rosser2019,trouille2019citizen}. 
A substantial portion of Zooniverse projects focus on wildlife occurrence, abundance, and behaviors, with volunteers classifying images from networks of remote automatically-triggered cameras, or ``camera traps.''
As these technologies become cheaper and more accessible, use of camera traps to monitor wildlife presence and abundance is increasingly popular in ecology \cite{rowcliffe2008,mccallum2013}.
Camera trap networks offer a cost effective way to collect rigorous research data with minimal human interference, while simultaneously creating a massive data extraction problem.

A typical camera trap network produces thousands (potentially millions) of images per deployment, but human participation remains a necessity for classifying wildlife in most images since computer vision is not yet reliable enough for this task \cite{Yousif_Yuan_Kays_He_2019, Willi_Pitman_Cardoso_Locke_Swanson_Boyer_Veldthuis_Fortson_2019, Anton_Hartley_Geldenhuis_Wittmer_2018}.
Deploying camera traps in natural settings also generates a high rate of ``empty'' images (also referred to as ``blanks'') with no animals present due to cameras being triggered by something other than wildlife like grasses moving in the wind or humans doing maintenance on the cameras.
While evaluating whether an image is empty of animals seems a simple judgment -- is there or is there not an animal present? -- a variety of characteristics of image subjects have potential to cause confusion for volunteers and reduce the levels of consensus on classifications \cite{ahuja_impact_2019}. 

The question of whether or not animals are present is only one of several judgments that volunteers must make in classifying camera trap images.
Better understanding how volunteers classify these and other potentially confusing images can help in quality control and implementation of machine learning to filter content so that fewer human judgments are required \cite{beck2018,wright2017}.
The volume of camera trap data and number of projects also continues to grow at a faster rate than volunteer recruitment to these projects, creating additional pressures to increase productivity and efficiency of classification processes.
As of June 2021, approximately 25\% of active projects on Zooniverse focused on camera trap classifications, and they consistently form a high proportion of the active projects on Zooniverse.

As with blank images, strong consensus on species identification can reduce the number of classifications needed for satisfactory data.
This is often the case for distinctive species such as elephant or giraffe where human classifications are at or near 100\% consensus. \cite{Swanson_Kosmala_Lintott_Packer_2016,Hsing_Bradley_Kent_Hill_Smith_Whittingham_Cokill_Crawley_Stephens_2018}.
However, lower consensus may result when images contain animals that are less common, have similar appearances, are interspersed with other species, or have visual obstructions in the foreground or background.
This can limit potential data uses beyond the project's initial goals, and the increased complexity of the identification task may impact volunteers' performance, interest, long-term engagement with the project, and the value that they derive from the experience.
The efficient use of volunteer time and effort is also an ethical concern in citizen science projects where participants are unpaid volunteers.

Introducing design-based interventions, such as offering different classification response options or clarifying the task instructions to reduce volunteer confusion, may help to improve consensus.
Such design changes need to be empirically grounded to help overcome project leaders' uncertainty before they are open to making changes to project design \cite{Law_Gajos_Wiggins_Gray_Williams_2017}.
Using machine learning to streamline camera trap image classification, for example, is an active area of research and development, but most work focuses on relatively simple images with only one animal to identify, and few algorithms are robust to variation in the landscapes where animals are present \cite{anton2018monitoring, Swanson_Kosmala_Lintott_Packer_2016}.
As we show in this study, these are the images for which volunteers are already efficient and reliable, and their performance issues related to landscape variations are limited and predictable.
Even in the future, when machine learning can handle naturalistic content in camera trap images, researchers are likely to continue to rely on volunteer classifications to develop initial training data for new corpuses and features of interest, making issues around consensus an ongoing concern.

In this study, we explored the impact of uncertainty on volunteer efficiency across six projects, asking: \textit{Which features of a subject, such as number or diversity of species, lead to volunteer uncertainty, contribute to low levels of consensus on image classifications, and impact volunteer efficiency?}

Uncertainty has been studied in a variety of areas, yet we know surprisingly little about uncertainty from the volunteer's perspective  \cite{Law_Gajos_Wiggins_Gray_Williams_2017}, or how this type of uncertainty may impact volunteer efficiency and overall consensus on subject classification.
Common sense suggests that uncertainty about presence or identification of animals could lead volunteers to spend more time on classification than necessary and may impact retention when volunteers are frequently unsure of the value of their work.
We focus on understanding uncertainty to maximize usable data while also addressing issues that may ultimately hamper project engagement and the volunteer experience.

This study makes three primary contributions: first, we identify five categories of image content \textit{complexity} which can contribute to volunteer uncertainty and impact classification efficiency, and which are likely to have parallels in other image classification tasks.
Second, our analysis of how these categories impact volunteers' efficiency in making classifications confirmed their validity.
Third, we developed a schema of image content \textit{characteristics and quality issues} that may impair consensus on classification tasks.
Our results suggest that volunteer uncertainty may have a substantial impact on efficiency. 
These contributions support recommendations for future research and design considerations to improve participant experience and data quality in crowdsourced image classification.

\section{Background}

This work is grounded in our current understanding of data quality and uncertainty as it relates to crowdsourced work in general and citizen science more specifically.
We also discuss image content as a catalyst for volunteer uncertainty, which in turn can affect classification efficiency and overall volunteer engagement.

\subsection{Data Quality in Crowd Consensus Tasks}

Improving data quality from crowdsourcing inevitably relies on reducing the classification (or labeling) noise found in large data sets to more acceptable and usable levels \cite{Tu_Yu_Domeniconi_Wang_Xiao_Guo_2020}. 
The most common method, either for ongoing analysis or more automated ML tasks, is often agreement through repeated labeling, i.e., crowd consensus \cite{Jung_Lease_2011,Sheng_Provost_Ipeirotis_2008}. 
Depending upon the requirement for the data set, consensus can take on several levels of sophistication ranging from simple majority voting, which assumes the bulk of human contributions are independent and of high quality, to more complex algorithms which assume all labelers are prone to biases of sensitivity and specificity \cite{Jung_Lease_2011, Sheshadri_Lease}.
Yet, in spite of its many manifestations and degrees of precision, researchers often lack a clear understanding of the various consensus methods available, and as a result, resort to the default of consensus by simple majority vote \cite{Sheshadri_Lease,Williams_Goh_Willis_Ellison_Brusuelas_Davis_Law}.
For the most part, this approach appears to be satisfactory, with only modest improvements made through other more complex statistical methods relying on confusion matrices to weight and filter contributions \cite{Jung_Lease_2011,Sheshadri_Lease}.

More sophisticated consensus methods are not without their weaknesses. 
Many of these methods are tested on artificially created data sets that lack input from human contributors \cite{Tu_Yu_Domeniconi_Wang_Xiao_Guo_2020, Sheng_Provost_Ipeirotis_2008}. 
This impacts the generalizability of the results to other projects and often these data sets are created with static difficulty levels for the classification task itself, which may not accurately reflect the complexity of real-world data \cite{Sheng_Provost_Ipeirotis_2008}. 
One of the reasons that task difficulty is generally limited to simple single subject classifications is that current consensus methods cannot accurately or efficiently analyze multiple-choice selection tasks; i.e., testing the classification accuracy of only one object (reported in the literature as a multiple-class selection task where a single subject can be classified multiple ways in one task) versus the classification of every object actually present in an image (reported as a multiple-choice selection task where classifying each subject is done congruently within a single task) \cite{Sheshadri_Lease, Tu_Yu_Domeniconi_Wang_Xiao_Guo_2020}.

Although there are work-arounds that mitigate the challenges of analyzing multiple-choice selection tasks such as transforming these tasks into simpler multiple-class or binary tasks so that algorithms such as ZenCrowd, Raykar (RY), and Dawid \& Skene (DS) can be used, this process takes additional time and resources to complete, and observable correlations among the classified objects within the image can be lost \cite{Sheshadri_Lease,Zhang_Sheng_Li_Wu_Wu_2017, Tu_Yu_Domeniconi_Wang_Xiao_Guo_2020}. 
More recently, there has been a push to provide consensus algorithms for multiple-choice classification tasks.
Methods such as MLCC, RAkEL-GLAD, and C-DS provide researchers with robust methods for analysis of multiple-choice data sets, but much like consensus algorithms for multiple-class data, little is known how the algorithms for multiple-choice data will fare in the ``wild'' since this research appears to be preliminary and largely untested \cite{Tu_Yu_Domeniconi_Wang_Xiao_Guo_2020}.

Crowdsourced image classification projects found on Amazon Mechanical Turk, Figure Eight, and other paid crowdsourcing sites are similar to citizen science classification in as much as they require humans to analyze and classify specific objects of interest contained in an image, such as identifying x-rays or microscopy images from a biomedical database \cite{karger2011iterative}.
Image classification in paid crowdsourcing typically relies on the same redundancy techniques as in citizen science, i.e., repeated labeling.
However, paid crowdsourcing platforms allow screening of workers for specific skills, resulting in fewer people needed for consensus, sometimes compensating just two workers with a third brought in when agreement cannot be reached between the first two \cite{de2014crowdsourcing}.
This phenomenon is also observed n citizen science projects as well, where images containing charismatic (e.g., giraffe) or common animals (e.g., domestic cat) may require just two or three classifications to reach consensus levels of greater than 97\% \cite{Anton_Hartley_Geldenhuis_Wittmer_2018, Swanson_Kosmala_Lintott_Packer_2016}. 

Yet, for every image that contains easily identifiable objects that fit nicely into specific classifications, there are countless other images that are ambiguous and full of noise \cite{Schaekermann_Goh_Larson_Law_2018}.
This noise can come from the quality of the image (blurry, blank, over-exposed, etc.) or from the content within the image itself, such as rare, multiple, or easily confused species, and often requires many more classifications from volunteers to reach consensus with satisfactory confidence levels for the intended research goals \cite{Swanson_Kosmala_Lintott_Packer_2016, Hsing_Bradley_Kent_Hill_Smith_Whittingham_Cokill_Crawley_Stephens_2018}.
In fact, Hsing et al. \cite{Hsing_Bradley_Kent_Hill_Smith_Whittingham_Cokill_Crawley_Stephens_2018} noted that mixed classifications for blank images, where something other than ``nothing here'' was chosen for an image devoid of wildlife, impacted consensus so much that confidence levels for that classification could not recover beyond 97.5\% despite multiple passes.
Even when an animal is present, the relative commonness of the species can impact the rate of classification accuracy and consensus. 
This has been observed with both false positive classifications of uncommon species as well as false negative classifications of more common species in a number of studies utilizing image classification of camera trap data \cite{Swanson_Kosmala_Lintott_Packer_2016, Steger_Butt_Hooten_2017}. 
Adding to the disparities identified in classifying simpler images, the accuracy and consensus rates for images containing multiple species creates an additional layer of complexity in image classification tasks, which remains too difficult for automation at this time \cite{Swanson_Kosmala_Lintott_Packer_2016, Willi_Pitman_Cardoso_Locke_Swanson_Boyer_Veldthuis_Fortson_2019}.
In particular, prior research has primarily reported on consensus at the level of the entire image, rather than at a more granular level focused on multiple objects within the image.

\subsection{Data Quality and Uncertainty in Citizen Science}

With the proliferation of camera trap imaging in conservation research, volunteers are increasingly important for subject classification of images provided to them online \cite{Clare_Townsend_Anhalt-Depies_Locke_Stenglein_Frett_Martin_Singh_Deelen_Zuckerberg_2019}.
Data quality of volunteer contributions to scientific research has been a topic of great interest in recent years, albeit with mixed reviews \cite{Law_Gajos_Wiggins_Gray_Williams_2017,Steger_Butt_Hooten_2017}
For example, in a review of 63 citizen science projects, comparison of citizen science data to gold standard or ground truth data was measured to have met minimum levels of accuracy in just 51\% - 62\% of the projects evaluated despite project reports to the contrary \cite{Aceves-Bueno_Adeleye_Feraud_Huang_Tao_Yang_Anderson_2017}. 
However, this evaluation was conducted with different goals and assessments of quality than the initial projects, and many projects independently evaluate their volunteers' work and find the data are of adequate (or better) quality for the project's intended purposes \cite{Willi_Pitman_Cardoso_Locke_Swanson_Boyer_Veldthuis_Fortson_2019, Hsing_Bradley_Kent_Hill_Smith_Whittingham_Cokill_Crawley_Stephens_2018}. 

Volunteer uncertainty offers a novel lens for understanding volunteer efficiency and engagement, but ``classification anxiety'' has been a focus of research surrounding volunteer motivation in online citizen science for a number of years \cite{Eveleigh_Jennett_Blandford_Brohan_Cox_2014}.
Such work has indicated that classification anxiety was a source of project drop-out due to volunteers feeling that the work submitted was not ``good enough'' for the project's scientific endeavors or the research was otherwise ``out of their league'' \cite{Eveleigh_Jennett_Blandford_Brohan_Cox_2014, Dayan_Gal_Segal_Shani_Cavalier}.
While we might infer that volunteer uncertainty was a factor in classification anxiety, prior work has not investigated the causes of classification anxiety \cite{Eveleigh_Jennett_Blandford_Brohan_Cox_2014, Segal_Gal_Simpson_Victoria_Homsy_Hartswood_Page_Jirotka_2015, Dayan_Gal_Segal_Shani_Cavalier}.
One study, however, looked at confidence levels of volunteer labelers \cite{Anton_Hartley_Geldenhuis_Wittmer_2018}, finding that volunteers' reported confidence levels in their classifications directly related to the level of accuracy of that classification.
In other words, as volunteers' confidence levels increased from ``unsure'' to ``reasonably confident'' to ``very confident,'' their classification accuracy improved as well \cite{Anton_Hartley_Geldenhuis_Wittmer_2018}.
Our work builds on this foundation, identifying potential sources of uncertainty that may impact classification accuracy at the individual level, and crowd consensus at the aggregate level.

\section{Methods}

We discuss relevant details of the study site and projects, and then briefly review key terminology used to describe the specifics of image classification in the Snapshot Safari projects, the data used for our analysis, and our mixed methods trace ethnography research strategy.

\subsection{Snapshot Safari and the Zooniverse Platform} \label{platform}

The Zooniverse platform supports online image analysis and transcription for scientific studies across several disciplines, has cumulatively involved millions of volunteers in authentic research experiences, and has supported over 100 studies since the platform launched.
Prior work in CSCW on the participation dynamics of Zooniverse projects has focused on newcomer movement, sustaining engagement, as well as volunteer learning \cite{jackson2018, Jackson_Crowston_Mugar_Osterlund_2016, Mugar_Osterlund_Hassman_Crowston_Jackson_2014}.

\begin{figure*}[h]
\includegraphics[width=\textwidth, height=8cm]{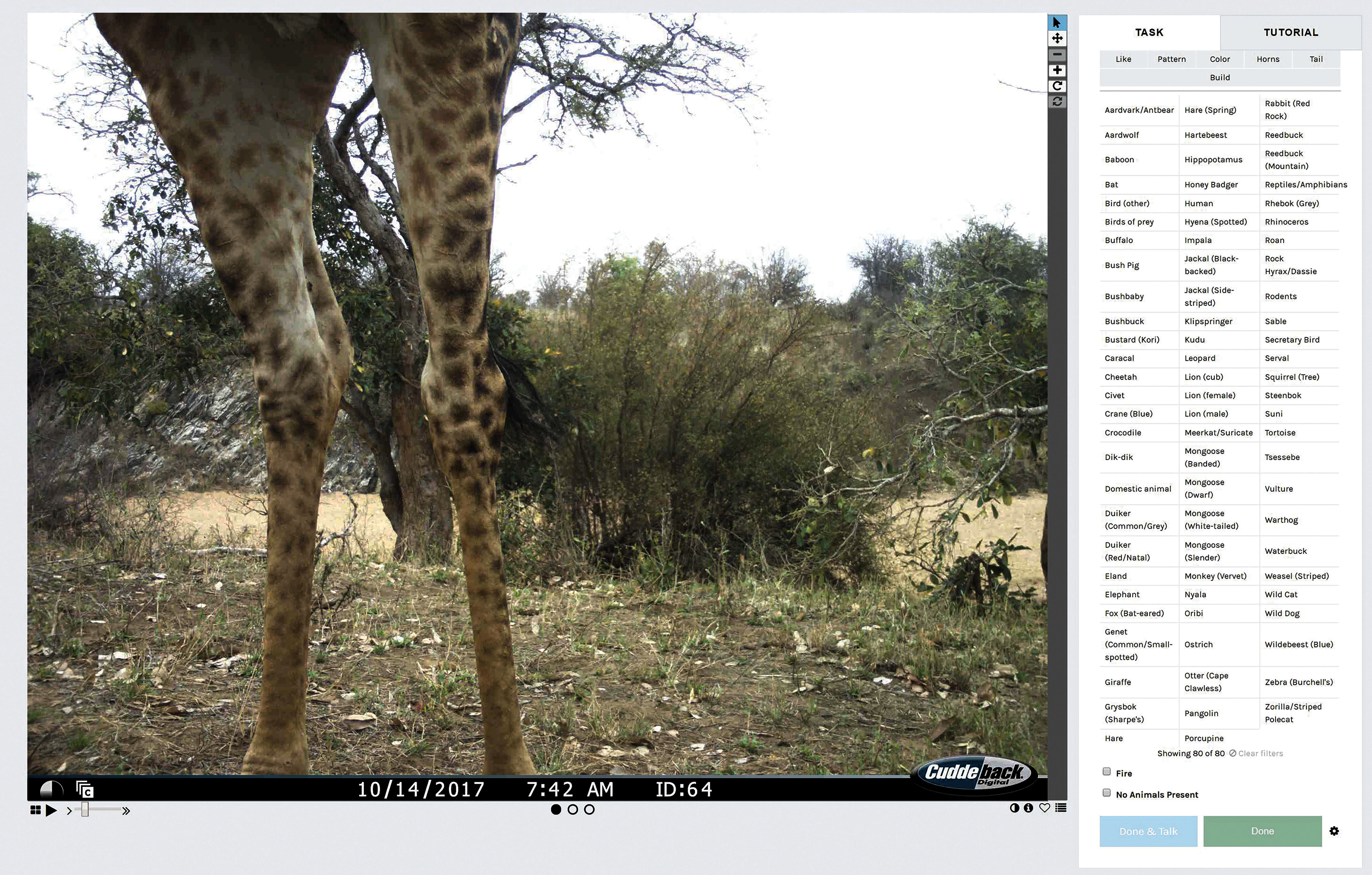}
\caption{The classification interface for Snapshot APNR, showing the subject to classify, image viewing tools, and response options. Other Snapshot Safari projects use an identical interface with species options suitable for their location.} \label{fig:APNR}
\end{figure*}

Zooniverse's Project Builder functionality launched in 2015 and provides a standardized set of tools for creating hosted online citizen science projects \cite{bowyer2015panoptes}, including the ``survey'' tool that is commonly used for camera trap image classification, as in the Snapshot Safari projects.
In this type of classification task, participants view either one or several (sequential) images from camera traps and identify the species or other content of interest by selecting the appropriate response from a grid of options (see Figure \ref{fig:APNR} for an example from Snapshot APNR.)
Standard tools include zoom, pan, rotate, reset view, filter, and invert image color (useful for night images).

Snapshot Safari is a collaborative network of camera trap grids in national parks and reserves throughout Africa, with each site generating thousands of images per location.
This study is based on data from six sites:
\begin{enumerate}
\item Associated Private Nature Reserves (APNR), on the boundary of Kruger National Park in South Africa
\item Gondwana, situated at the Southern Cape of South Africa
\item Grumeti, adjacent to Serengeti National Park in Tanzania
\item Mariri, in Niassa National Reserve in Northern Mozambique
\item Ruaha, located at Ruaha National Park in Tanzania 
\item Serengeti, at Serengeti National Park in Tanzania
\end{enumerate}

In prior studies of volunteer classification accuracy for Snapshot Serengeti, images for the gold standard corpus were selected at random, including images with no animals and potentially confusing content such as partial animals, e.g., the hindquarter of an ungulate.
The analysis of volunteer accuracy included 4,149 gold-standard images in a consensus data set to assess the data quality of 28,000 registered and nearly 40,000 unregistered volunteer participants who contributed 10.8 million classifications for 1.2 million images \cite{swanson2015snapshot}.
Of the 1.2 million image subjects, roughly one-third contained animals while the remaining subjects were classified as blanks.
Volunteers were 96.6\% accurate for species identification and 90\% accurate for species population counts \cite{swanson2015snapshot}.
Subsequent research has demonstrated similar accuracy levels for species identification, however findings to date have been limited to single-species images given the complex statistical analysis required of multiple-species images \cite{Swanson_Kosmala_Lintott_Packer_2016, Willi_Pitman_Cardoso_Locke_Swanson_Boyer_Veldthuis_Fortson_2019}.

In our research, Snapshot Safari projects offered a unique opportunity to understand the role of uncertainty in volunteer classification behavior, consensus, and classification efficiency, as the natural occurrence of blanks and presence of multiple species varies substantially across the six projects.
The camera trap networks installed in each park use similar deployment strategies for camera placement.
While each project features images from a different park with a different set of wildlife and habitat features, the projects are otherwise identical in their interface configuration, tutorial content, task requirements, and research team management, creating an ideal opportunity for cross-project comparison.
In the online classification interfaces, the only differences are species featured in tutorials and response options for species occurring in different parks (53--80 species.)
For each project, volunteers view an image and select a response from the specified species (or ``none''), and once the species are identified, they are asked to identify how many are present, their behaviors (e.g., moving, grazing, or eating), and presence of young.

Any variation in performance across projects would be primarily attributable to either content or the self-selected volunteer populations participating in each project (volunteer demographics are not currently collected).

Variability in the park habitats and management does have clear impacts on the content of images which are not attributable to variations in volunteer populations.
For example, Snapshot Ruaha is a grassland and movement of grasses in the wind frequently triggers motion-sensitive cameras, but in Snapshot Serengeti, which is also a grassland, stewards trim the grasses in front of cameras to reduce the incidence of false positive image captures.
By contrast, Snapshot APNR documents multiple ecosystems surrounding Kruger National Park and does not have this issue.
In addition, Snapshot Mariri's local partners pre-filtered their data to remove blank images and streamline classification, resulting in an artificially low rate of blank images for that project.

\subsection{Terminology}

We now introduce key terminology to support description of our data sources and analysis methods.

A \textit{project} is a collection of one or more workflows branded as belonging to a single campaign, typically with a single large image corpus from a given location or instrument.

A \textit{workflow} is a series of tasks created with the Project Builder tool that are applied to a specific corpus. 
There can be many workflows for any one project, but participants complete classification tasks within only one workflow at a time. 

A \textit{subject} is defined as the contents of any one camera capture presented to volunteers for classification. 
Each camera capture may include several components, like a series of three photos which make up a single subject, or may present non-photographic content in other projects.

A \textit{target} is the object of interest to be classified in the subject. 
For Snapshot Safari, the targets are wild animals found in various regions in Africa.
We also describe these targets as ``species'' since the core classification task is identification of the animals to the level of species in the taxonomic hierarchy.
Each subject may contain multiple targets

A \textit{classification} is the decision resulting from the process of individual participants identifying the target of a subject within the project's classification interface. 
In Snapshot Safari projects, these tasks involve matching the corresponding animal species found on the option grid to the targets in the subject.

\textit{Low consensus} subjects are those for which aggregate classifications show less than 50\% agreement on whether it is blank, if an animal is present, or which type of species is present.
For example, a subject with 14 classifications of blank and 16 classifications of any other option besides blank is considered low consensus.
As defined by the Snapshot Safari research team, subjects with less than 75\% consensus on the species identification are not used for research. 
 
We examined both simple majority and research team thresholds for our assessment. 
We chose the lower consensus threshold to sample content with greater ambiguity, since subjects with low consensus based on simple majority will also fail to meet the reliability threshold for species identification.

\textit{High consensus} subjects are those for which the total responses met or exceeded the 75\% consensus threshold defined by Snapshot Safari. 
According to Hines et al. \cite{hines2015aggregating}, an algorithm retires blank subjects with high consensus when the first five classifications are blank or when there are ten total classifications of blank.

\textit{Efficiency} is the length of time it takes a volunteer to classify a subject, measured in seconds \cite{jackson2016way}.
When volunteers are able to accurately classify a subject more quickly, they can classify more subjects and generate more data for scientific research.

\subsection{Data}

The Snapshot Safari science team provided access to system-generated download files of subjects and classifications made by volunteers.
We also accessed the ``Talk'' comments made by volunteers on specific subjects for content analysis.
\footnote{Since these data are also ecology research data for the Snapshot Safari team, who provided access for this work, we do not currently have permission to release it as open data. Interested parties are encouraged to contact the Snapshot Safari team for data access or the authors to discuss collaboration opportunities.}
Classification data included timestamps, subject IDs, (anonymous) user IDs, browser strings, and the classification decisions made by volunteers.
We loaded these files into a PostgreSQL database on a local high-performance computing cluster for analysis. 
The earliest date in our analysis was July 24, 2017 and the last date was June 26, 2019. 
The duration of data for each project in our analysis, listed in Table \ref{table:datasummary}. 

The data contained a total of 5,826,254 classifications on 644,395 subjects across the six projects; we used the entirety of the data set, aside from outlier filtering as described below.
Content in Talk dated after August 2018 was also reviewed to provide context for statistical analyses, and included comments from volunteers and moderators in the APNR, Grumeti, Ruaha, and Serengeti projects.

\begin{table}
\centering
\caption{Descriptive Statistics for Snapshot Safari Projects}
\resizebox{\linewidth}{!}{%
\begin{tabular}{lccccccc}
\multicolumn{1}{c}{\multirow{6}{*}{\textbf{Project}}} &
  \multirow{4}{*}{\textbf{\begin{tabular}[c]{@{}c@{}}Total \\ Subjects\end{tabular}}} &
  \multirow{4}{*}{\textbf{\begin{tabular}[c]{@{}c@{}}Total \\ Classifications\end{tabular}}} &
  \multirow{4}{*}{\textbf{\begin{tabular}[c]{@{}c@{}}Mean \\ classification \\ time per \\ subject (sec) \end{tabular}}} &
  \multirow{4}{*}{\textbf{\begin{tabular}[c]{@{}c@{}}Median \\ classification \\ time per \\ subject (sec)\end{tabular}}} &
  \multirow{4}{*}{\textbf{\begin{tabular}[c]{@{}c@{}}Unique \\ users\end{tabular}}} &
  \multirow{4}{*}{\textbf{\begin{tabular}[c]{@{}c@{}}Species \\ Options\end{tabular}}} &
  \multirow{4}{*}{\textbf{\begin{tabular}[c]{@{}c@{}}Days \\ in \\ Analysis\end{tabular}}} \\
\multicolumn{1}{c}{} &         &           &       &    &        &    &     \\
\multicolumn{1}{c}{} &         &           &       &    &        &    &     \\
\multicolumn{1}{c}{} &         &           &       &    &        &    &     \\
\multicolumn{1}{c}{} &         &           &       &    &        &    &     \\
\multicolumn{1}{c}{} &         &           &       &    &        &    &     \\
APNR                 & 41,302  & 411,749   & 19.47 & 13 & 3,142  & 80 & 505 \\
Gondwana             & 5,089   & 109,986   & 10.39 & 4  & 1,750  & 53 & 532 \\
Grumeti              & 43,109  & 506,799   & 26.9  & 18 & 8,259  & 62 & 554 \\
Mariri               & 14,372  & 288,945   & 30.44 & 20 & 3,505  & 63 & 550 \\
Ruaha                & 351,228 & 2,283,839 & 7.74  & 5  & 5,066  & 63 & 559 \\
Serengeti            & 189,295 & 2,224,936 & 17.06 & 9  & 17,723 & 57 & 702
\end{tabular}
}
\label{datasummary}
\label{table:datasummary}
\end{table}

Additional project-level details are provided in Table \ref{table:datasummary}, since the number of subjects, classifications, and unique users varied substantially across the projects.
Snapshot Gondwana had a far smaller subject set than the other projects, while Snapshot Serengeti has had the most data classified.
The variance in the number of classifications per subject in each project reflects different retirement rules for removing subjects from the workflows over the period of time under study.
The number of classifications required for retirement was initially much higher (approximately 20 classifications per subject), until analysis showed that consensus could be reached with about half as many judgments \cite{Swanson_Kosmala_Lintott_Packer_2016}.
In addition, we report mean and median time (excluding outliers) to classify subjects, number of species listed in response options, and days of project activity analyzed. 
Outliers removed from analysis were classifications that either showed a zero or ``negative'' completion time (due to occasional logging errors) or inordinately long duration for a single classification.
In these cases, volunteers had loaded the subject for classification but did not complete the task in the same session, leaving a browser window open for hours or even days, and later completing the classification in a separate session.
For calculations related to classification time, the top 5\% of subject classification duration were removed.

\subsection{Classification Consensus Categories} \label{categorization}

This work initially focused on low consensus in blank image classification.
A trace ethnography strategy of progressive data exploration, along with our content analysis described below, revealed several emergent categories of consensus on classifications.
We were inspired to look more closely at the impact of multiple species on consensus when prior research reported nearly 100\% agreement on subjects containing giraffes but in our initial analyses, subjects classified as containing giraffes did not have high consensus.
After careful inspection of the data, we found that the volunteers were indeed in very strong agreement about giraffes, but it was uncertainty about the \textit{other} animals present alongside giraffes that led to reduced consensus at the subject level.
Essentially, for some subjects, low consensus appeared to have more to do with the number of species present, rather than uncertainty about a given species such as giraffes.
We then derived several categories of subjects based on extended examination of the data, and subsequently conducted statistical tests to confirm their independence, reported in the results.

\begin{itemize}
  \item \textit{Blank}, classified as blank (100\% agreement).
  \item \textit{Single Species}, classified as a single species (100\% agreement). 
  \item \textit{Blank or Single Species}, classified as either blank or as a single species. 
  \item \textit{Two Species}, classified as either of two species, but not blank.
  \item \textit{Multiple Species}, classified as more than two species and/or blank.
\end{itemize}

\subsection{Analysis}

We applied mixed methods analysis, with qualitative content analysis, descriptive statistics, and group comparison among the classification consensus categories described above.

\subsubsection{Statistical Analysis}

Our statistical analyses used a combination of SQL queries and the R statistical software to retrieve, classify into groups, and analyze classification data.

The focus was to understand the differences in volunteer effort (measured as time in seconds) among different classification groups and how these might reflect uncertainty affecting volunteer performance.

To statistically check the normality assumption, the Shapiro Test was performed; results indicated that not all groups are normally distributed.
Therefore, the Kruskal–Wallis test was performed to compare the differences in time efficiency among the groups.
To further assess the differences between groups, the post-hoc Dunn test was performed.

\begin{figure*}[h]
\includegraphics[width=\textwidth, height=10cm]{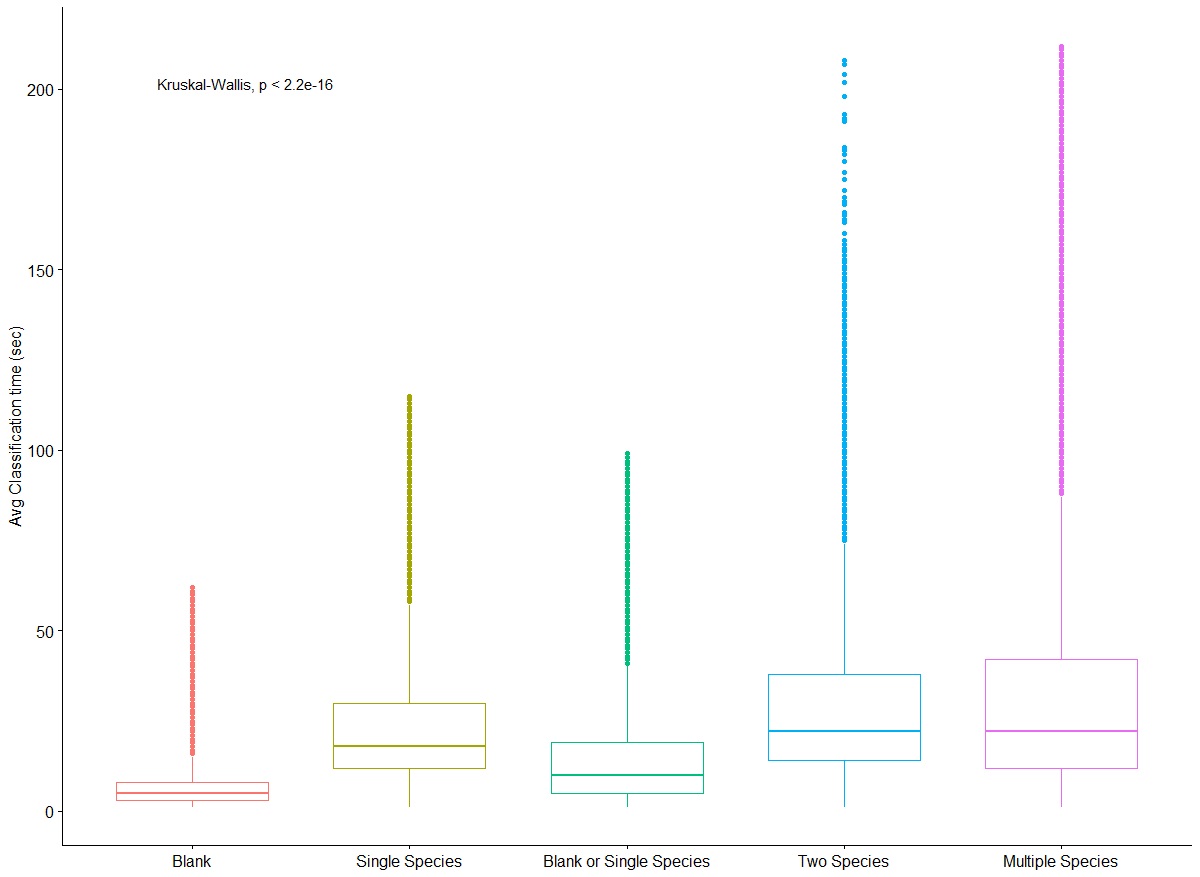}
\caption{Box Plot for all the groups} 
\label{fig:boxplot}
\end{figure*}

\subsubsection{Content Analysis}

Our analysis of the subject content followed an inductive content analysis process \cite{neuendorf}, with two independent coders evaluating a random selection of low consensus subjects from all six projects.
We conducted two rounds of content analysis: in the first round, 92 low consensus subjects were selected from across the six projects and were reviewed to identify categories of potential sources of confusion for classification.
All subjects with at least one blank classification were extracted and simple consensus on the presence of blank or not was calculated. 
Low consensus subjects that also had related Talk comments (n = 1,755) were selected as a sampling frame, and 92 subjects were randomly sampled for content analysis.

Two coders then met to discuss judgments of whether or not an animal was present for the initial sample. 
An inductive content analysis schema was developed through this process to characterize the potential sources of confusion for Zooniverse volunteers.
Many of the sources of confusion were readily apparent to the researchers, but reviewing the corresponding ``Talk'' data to elicit additional clues assisted in refining the coding schema. 
Each coder first determined whether they believed there was an animal present and then identified any features of the subjects that could have caused confusion or disagreement in volunteers' responses.
For example, some subjects showed animals present on a distant horizon which were only detectable after zooming in on the image.
Other subjects had characteristics such as low light or partial animals in the frame.

After completing independent coding, both coders met to reconcile differences in their assessments (arguing to consensus) and further examined image subjects for which their evaluations did not conform to the simple majority consensus of the volunteers.
The initial rate of agreement was 68\%, and the coders conducted detailed evaluation of 29 images for variance between their evaluations and volunteers' responses.
After resolving coding process inconsistencies and arguing to consensus, the level of agreement was 97\% and the coders had agreed to disagree on three subjects which had features that project instructions were ambiguous on how to handle.
For the second round of analysis, we purposively sampled 332 low consensus subjects, and one coder applied the schema to all subjects to identify the frequency with which confusing subject features occurred.

\section{Findings}

Our results focus on content characteristics impacting volunteers' responses and consensus levels across projects.

\subsection{Group Differences in Classification Consensus Categories}

The Kruskal–Wallis test indicated (chi-square = 2343614, df 4, p-value < 2.2e-16) that there were significant differences in the amount of time that volunteers spent on classification based on our categorization of subjects (discussed in Section \ref{categorization}).
The Dunn multiple comparison among groups results were also significant, indicating that the classification consensus category groups vary significantly in terms of the amount of time volunteers spent on each type of subject.

\begin{table}
\centering
\caption{Mean classification time in seconds for each Classification Consensus category}
\resizebox{\linewidth}{!}{%
\begin{tabular}{ccccccccccccccccc}
\multirow{5}{*}{\textbf{Project}} & \multirow{5}{*}{\textbf{\begin{tabular}[c]{@{}c@{}}Total \\ Subjects\end{tabular}}} & \multicolumn{2}{c}{\multirow{4}{*}{\textbf{Blank}}} & \multirow{5}{*}{\textbf{\begin{tabular}[c]{@{}c@{}}Mean \\ Time \\ (sec)\end{tabular}}} & \multicolumn{2}{c}{\multirow{4}{*}{\textbf{\begin{tabular}[c]{@{}c@{}}Single\\ Species\end{tabular}}}} & \multirow{5}{*}{\textbf{\begin{tabular}[c]{@{}c@{}}Mean \\ Time \\ (sec)\end{tabular}}} & \multicolumn{2}{c}{\multirow{4}{*}{\textbf{\begin{tabular}[c]{@{}c@{}}Blank or\\ Single \\ Species\end{tabular}}}} & \multirow{5}{*}{\textbf{\begin{tabular}[c]{@{}c@{}}Mean \\ Time \\ (sec)\end{tabular}}} & \multicolumn{2}{c}{\multirow{4}{*}{\textbf{\begin{tabular}[c]{@{}c@{}}Two \\ Species\end{tabular}}}} & \multirow{5}{*}{\textbf{\begin{tabular}[c]{@{}c@{}}Mean \\ Time \\ (sec)\end{tabular}}} & \multicolumn{2}{c}{\multirow{4}{*}{\textbf{\begin{tabular}[c]{@{}c@{}}Multiple \\ Species\end{tabular}}}} & \multirow{5}{*}{\textbf{\begin{tabular}[c]{@{}c@{}}Mean \\ Time \\ (sec)\end{tabular}}} \\
 &  & \multicolumn{2}{c}{} &  & \multicolumn{2}{c}{} &  & \multicolumn{2}{c}{} &  & \multicolumn{2}{c}{} &  & \multicolumn{2}{c}{} &  \\
 &  & \multicolumn{2}{c}{} &  & \multicolumn{2}{c}{} &  & \multicolumn{2}{c}{} &  & \multicolumn{2}{c}{} &  & \multicolumn{2}{c}{} &  \\
 &  & \multicolumn{2}{c}{} &  & \multicolumn{2}{c}{} &  & \multicolumn{2}{c}{} &  & \multicolumn{2}{c}{} &  & \multicolumn{2}{c}{} &  \\
 &  & \textbf{n} & \textbf{\%} &  & \textbf{n} & \textbf{\%} &  & \textbf{n} & \textbf{\%} &  & \textbf{n} & \textbf{\%} &  & \textbf{n} & \textbf{\%} &  \\
APNR & 41,302 & 5,630 & 14\% & 9.77 & 22,746 & 55\% & 19.93 & 3,513 & 9\% & 13.51 & 4,400 & 11\% & 28.29 & 5,013 & 12\% & 29.66 \\
Gondwana & 5,089 & 3,420 & 67\% & 4.22 & 181 & 4\% & 25.75 & 958 & 19\% & 7.14 & 79 & 2\% & 44.63 & 451 & 9\% & 34.1 \\
Grumeti & 43,109 & 6,559 & 15\% & 11.77 & 6,786 & 16\% & 26.62 & 8,768 & 20\% & 20.55 & 4,537 & 11\% & 31.61 & 16,459 & 38\% & 32.05 \\
Mariri & 14,372 & 149 & 1\% & 12.42 & 3,524 & 25\% & 21.19 & 1,301 & 9\% & 20.55 & 2,604 & 18\% & 27.32 & 6,794 & 47\% & 38.67 \\
Ruaha & 351,228 & 311,643 & 89\% & 5.48 & 13,717 & 4\% & 29.03 & 15,344 & 4\% & 10.59 & 5,482 & 2\% & 35.2 & 5,042 & 1\% & 33.69 \\
Serengeti & 189,295 & 109,005 & 58\% & 8.32 & 8,354 & 4\% & 27.62 & 27,878 & 15\% & 14.91 & 7,680 & 4\% & 32.54 & 36,378 & 19\% & 31.1
\end{tabular}
}
\label{classificationtime}
\label{table:classificationtime}
\end{table}

Further, table \ref{table:consensus} shows that as the number of species classified in a subject increases, the level of consensus among volunteers decreases.
That is, when more animals appeared in the subject, overall subject-level consensus was lower.
However, we observed that there is usually consensus on at least one species in these multi-target subjects.
This means that there may be 100\% agreement on one target, such as giraffe, while a second or third target might have dramatically lower consensus.
As a result, when using simple subject-level consensus measures with the 75\% agreement threshold, the entire subject appears to have low consensus.
In actuality, there is likely one high consensus target and one or more low consensus targets.
For example, in the Blank or Single Species category across all six projects, 83\% to 97\% had high consensus, but for subjects that fell into the Multiple Species category, the overall agreement was reduced to between 37\% and 55\%.
Evaluating the Multiple Species category with a 50\% consensus threshold showed agreement levels of 83\% to 90\%, evidence of strong agreement on one species out of several.

\begin{table}
\centering
\caption{Consensus on Multiple Species}
\resizebox{\linewidth}{!}{%
\begin{tabular}{l|ccc|ccc|ccccc}
\multicolumn{1}{c}{\multirow{5}{*}{\textbf{Project}}} & \multirow{6}{*}{\textbf{\begin{tabular}[c]{@{}c@{}}Blank \\ or \\ Single\\ Species\end{tabular}}} & \multicolumn{2}{c}{\multirow{5}{*}{\textbf{\begin{tabular}[c]{@{}c@{}}\textgreater{}= 75\%   \\ Agreement\end{tabular}}}} & \multirow{6}{*}{\textbf{\begin{tabular}[c]{@{}c@{}}Two \\ Species\end{tabular}}} & \multicolumn{2}{c}{\multirow{5}{*}{\textbf{\begin{tabular}[c]{@{}c@{}}\textgreater{}= 75\%   \\ Agreement\end{tabular}}}} & \multirow{6}{*}{\textbf{\begin{tabular}[c]{@{}c@{}}Multiple \\ Species\end{tabular}}} & \multicolumn{2}{c}{\multirow{5}{*}{\textbf{\begin{tabular}[c]{@{}c@{}}\textgreater{}= 75\%   \\ Agreement\end{tabular}}}} & \multicolumn{2}{c}{\multirow{5}{*}{\textbf{\begin{tabular}[c]{@{}c@{}}\textgreater{}= 50\%   \\ Agreement\end{tabular}}}} \\
\multicolumn{1}{c}{} &  & \multicolumn{2}{c}{} &  & \multicolumn{2}{c}{} &  & \multicolumn{2}{c}{} & \multicolumn{2}{c}{} \\
\multicolumn{1}{c}{} &  & \multicolumn{2}{c}{} &  & \multicolumn{2}{c}{} &  & \multicolumn{2}{c}{} & \multicolumn{2}{c}{} \\
\multicolumn{1}{c}{} &  & \multicolumn{2}{c}{} &  & \multicolumn{2}{c}{} &  & \multicolumn{2}{c}{} & \multicolumn{2}{c}{} \\
\multicolumn{1}{c}{} &  & \multicolumn{2}{c}{} &  & \multicolumn{2}{c}{} &  & \multicolumn{2}{c}{} & \multicolumn{2}{c}{} \\
\multicolumn{1}{c}{\textbf{}} &  & \textbf{n} & \textbf{\%} &  & \textbf{n} & \textbf{\%} &  & \textbf{n} & \textbf{\%} & \textbf{n} & \textbf{\%} \\
APNR & 3,513 & 3,229 & 92\% & 4,400 & 2,742 & 62\% & 5,013 & 2,568 & 51\% & 4,305 & 86\% \\
Gondwana & 958 & 926 & 97\% & 79 & 71 & 90\% & 451 & 205 & 45\% & 373 & 83\% \\
Grumeti & 8,768 & 7,616 & 87\% & 4,537 & 4,311 & 95\% & 16,459 & 7,360 & 45\% & 14,756 & 90\% \\
Mariri & 1,301 & 1,280 & 98\% & 2,604 & 2,433 & 93\% & 6,794 & 3,739 & 55\% & 5,700 & 84\% \\
Ruaha & 15,344 & 12,779 & 83\% & 5,482 & 4,181 & 76\% & 5,042 & 1,888 & 37\% & 4,295 & 85\% \\
Serengeti & 27,878 & 24,200 & 87\% & 7,680 & 6,753 & 88\% & 36,378 & 13,569 & 37\% & 30,693 & 84\%
\end{tabular}
}
\label{consensus}
\label{table:consensus}
\end{table}

\subsection{Characteristics of Low Consensus Images}

We also applied qualitative analysis techniques to identify sources of uncertainty for volunteers.

The two primary categories for potential sources of confusion that emerged were subject quality and target characteristics.
Subject quality issues related to attributes of the image presented, while target characteristics related to variations around how the animal was pictured in its environment.
There were also ten subjects with ``other'' issues, unique to the specific content, omitted from the reporting.

Subject quality issues included blurry or low resolution images, over- and under-exposure, and low contrast between background and animal.
Subject quality issues were identified in 36\% of low consensus subjects evaluated by our coders, with image exposure (61\%) and contrast (30\%) as the most common problems.
While these results are unsurprising, they confirm that managing subject quality could have potential to improve consensus levels.

The target characteristics that were associated with low consensus were more nuanced: distance between the target animal and the camera, obstruction or partial capture of the animal in the image frame, and rare animals that did not appear in the response options.
The overall occurrence of problematic target characteristics was 64\% of the low consensus subjects we evaluated, with most common being lack of a response option for the species (33\%), distant animals (26\%), and partial or obstructed views of the animal (23\%).
Again, these were not known or assumed issues prior to this analysis, so confirmation of the specifics encourages design interventions for the particulars of these challenges, rather than on the basis of assumptions.

\begin{table}[h]
\caption{Potential sources of confusion for image analysis.}
\begin{tabular}{l l l l}
 & \textbf{\begin{tabular}[c]{@{}l@{}}Subjects \\ Reviewed\end{tabular}} & \textbf{\begin{tabular}[c]{@{}l@{}}\% of \\ Issue\end{tabular}} & \textbf{\begin{tabular}[c]{@{}l@{}}\% of \\ Total \\ (322)\end{tabular}} \\ 
\multicolumn{4}{l}{\textbf{Issue: Subject Quality}} \\
Over/Under-exposed & 71 & 61\% & 22\% \\ \hline
Low Contrast & 35 & 30\% & 11\% \\ \hline
Blurry & 10 & 9\% & 3\% \\ \hline
\textit{Subject Quality Total} & \textit{116} &  & \textit{36\%} \\ 
& & & \\
\multicolumn{4}{l}{\textbf{Issue: Target Characteristics}} \\ \hline
No response option & 69 & 33\% & 21\% \\ \hline
Too far & 54 & 26\% & 17\% \\ \hline
Partial/obstructed figure & 48 & 23\% & 15\% \\ \hline
Too close & 22 & 11\% & 7\% \\ \hline
Small target & 13 & 6\% & 4\% \\ \hline
\textit{Target Characteristics Total} & \textit{206} &  & \textit{64\%} \\ 
\end{tabular}
\label{tab:contentanalysis}
\end{table}

Overall, the two most common issues observed were lack of a response option for the species and under- or over-exposed images due to the angle of the sun or range of camera flash in night images, each accounting for 22\% of potential causes of confusion.
The next most common issues had to do with the distance of the animal from the camera (17\%) and images containing only a part of an animal or an obstruction (15\%).
Although some animals are generally easy to identify from just a haunch or legs, like zebras and giraffes (as shown in Figure \ref{fig:APNR}), ungulates such as gazelle species can be difficult to distinguish without a view of the entire animal.

\subsection{Uncertainty Expressed in Volunteers' Comments}

To further assess the impact of lack of response options, we filtered all Talk data from the six projects for keywords related to uncertainty (``uncertain'', ``unsure'', ``don't know'', ``nothing here'', ``not sure'', ``wasn't sure'', ``isn't sure'') to evaluate cases where volunteers specifically expressed uncertainty about the species present or their classification of it. 
We found that expressions of uncertainty were made by volunteers at all levels of experience, from novice volunteers classifying their first image to an expert volunteer who had classified 41,667 images.
Out of the 32,795 Talk comments filtered for this analysis, just over 2\% (456 comments) included expressions of uncertainty.
Excluding moderator responses (88), volunteers made 367 comments.
We assume, however, that the actual incidence of such experiences is much higher, and only a small proportion of volunteers are willing to comment on uncertainty, since only 8\% of unique users across the Snapshot Safari projects made any Talk comments regardless of topic.
Comments on uncertainty were made by 223 unique users, representing about 16\% of all volunteers who made any Talk comments (volunteers must be logged in to make comments), indicating that uncertainty is likely a common experience among volunteers.
About 30\% (113) of comments expressing uncertainty were focused specifically on species identification.

For example, one volunteer had classified a subject as blank, although there was clearly an animal in the image, and added it to a personal list using a platform feature called ``Collections,'' which makes it easy to revisit (but not reclassify) a specific subject.
The volunteer later saw others' Talk comments about the subject and added the comment,
\textit{``Yes! my `nothing here' is really a \#pangolin.''}
Notably, pangolins are endangered species that rarely appear in the corpus, so they are not listed as a response option.
This volunteer also demonstrated awareness of a convention among Zooniverse volunteers to use hashtags for identifying interesting and unusual features in subjects, including labeling species not listed among the response options, suggesting that they were not a novice.
Another volunteer mentioned their uncertainty at the time of rendering the classification, which showed that they had followed the instructions to make their best guess:
\textit{``Big ? on this one.  Called it a wildebeast [sic] because its an easy go to, but not sure its right.  Someone could probably identify based on the legs but not me.''}
This volunteer admits uncertainty and using a ``default'' species choice for classifying an ungulate rather than marking it as blank; wildebeests are quite common in the corpus, and therefore a reasonable guess.
These examples indicate that volunteers do not make these decisions lightly, and apply multiple specific strategies when classifying content that evokes uncertainty.

\section{Discussion}

Our results indicate opportunities for further research on the impact of subject content on crowd classifications, methods for assessing consensus on complex content, and potential design interventions to reduce task confusion.

\subsection{Impact of Subject Content on Crowd Classifications}

One readily apparent observation from our results is that certain categories of subjects are faster to classify than others.
Unanimously Blank subjects are fastest, followed by Blank or Single Species and then Single Species subjects.
Blank subjects take less classification time because no additional details must be provided, and when the number of species in the subject increases, the time to classify increases, since each species in the image must be identified and its demographics provided.
Durations for Blank or Single Species fall neatly between Blank subjects and Single Species subjects due to some volunteers not detecting the animal and others responding with ``blank'' for lack of confidence in their classification, but there is agreement on one species by all others.

The more complex subjects are those with Two Species and Multiple Species.
The longer classification times reflect the additional effort for volunteers deciding between two species when the target is apparent but its identity is not, if there are multiple types of animals present, or the animals present are more ambiguous and difficult to identify.

For subjects unanimously classified as blank, mean classification times were even lower when blank images were more common in the data set (Gondwana, Ruaha, Serengeti).
When blank subjects were uncommon (Mariri, Grumeti, APNR), volunteers took more time to classify these subjects, likely because they were accustomed to seeing animals present and therefore made additional effort to detect animals prior to rendering a decision. 
Gondwana shows slightly different trends than other projects; this is likely because the smaller corpus had very few subjects with any animals present at all.
Volunteers more quickly labeled the Blank subjects and the Blank or Single Species subjects (most of which are, in fact, blank), which together made up 86\% of the Gondwana corpus.
Similar patterns were observed in Ruaha with respect to time on task for the same two categories of subjects, which are 93\% of the corpus.
Both of these projects have lower mean and median classification times per subject than the others, reflecting the influence of the corpus content on  classification efficiency.
Gondwana volunteers also spent more time on Two Species subjects than Multiple Species, suggesting that additional effort was required to disambiguate similar-looking species when they are infrequently observed, and the occurrence of these subjects was very low.
These results are consistent with our content analysis findings: some images are more difficult to classify than others, whether or not they include animals.

\subsection{Sources of Uncertainty in Classification}

Our content analysis revealed several sources of uncertainty that may reduce consensus, including the complexity of the content, image quality, and issues related to interface specifics.

\subsubsection{Classification Consensus Categories and Content Complexity}

Our analysis indicates that complexity in subject content, as reflected in the classification consensus categories, is a notable source of uncertainty for volunteers. 
This uncertainty may lead to classification anxiety, which has been shown to negatively influence volunteer engagement, something most projects relying on crowdsourcing try to avoid \cite{Eveleigh_Jennett_Blandford_Brohan_Cox_2014, Dayan_Gal_Segal_Shani_Cavalier}.

We saw that even for the category with lowest consensus levels, Multiple Species, there was often high consensus on a single species, as reported in table \ref{consensus}.
This highlights an opportunity to extract reliable data from complex subjects which might otherwise be discarded due to low agreement with simplistic subject-level measures of consensus, delivering more value from volunteers' time investment.
Moreover, the results suggest that different conceptualizations and measures of consensus may be appropriate for different types of content, and the aggregate responses offer a way to identify content that needs different handling when these categories cannot be determined \textit{a priori}.
We were unable to locate prior work that offers analytical strategies or appropriate consensus measures for maximizing the usable data for this type of classification task, which indicates an opportunity for future work.

In practice, applying the same rules to all subjects, for both retirement and calculating agreement statistics, may result in unrecoverable data quality issues or eliminate the potential for future analysis of relationships among species.
Dynamic retirement rules and attention to how aggregate classifications signal content complexity in post-classification data analysis are potential strategies for mitigating these challenges.
Similar strategies are likely to work well for other citizen science and crowdsourcing projects where the subject content varies in complexity.

\subsubsection{Image Quality Issues}

The most common issue undermining consensus in otherwise simple subjects was overexposure or underexposure of the image.
Although there is little to be done about this, as it is an inherent challenge in camera trap image capture, response options could be adjusted.
In a prior study on Zooniverse, Anton et al. found that when volunteers were offered an option to indicate uncertainty, the use of this option correlated strongly with image quality problems \cite{anton2018monitoring}.
One potential design intervention, therefore, would be to offer a specific interface option to indicate that the subject is a poor quality image, in addition to the primary species classification.
This type of intervention has not previously been directly tested or evaluated, in part because of project teams' concerns that volunteers would see this as an ``easy out'' and put less effort into species identification if they can blame image quality for lack of accuracy.

Our results indicate otherwise, since images that included animals that were unidentifiable due to image quality were frequently marked blank, and Talk comments showed that volunteers were dissatisfied with having no option to directly indicate that there was an animal present but that it could not be classified due to image quality problems.
If implemented and determined reliable, this indicator could be used for adjusting retirement rules to remove an image as unclassifiable after multiple volunteers mark it as poor quality, improving overall efficiency of the classification workflow.
This would also likely improve the volunteers' experience, given their expressions of uncertainty in the Talk comments, taking full advantage of their ability to classify content.
The prior results from Anton et al. \cite{anton2018monitoring} suggest this intervention is worth testing with an experimental design to conclusively evaluate impacts on data quality.

\subsubsection{Identifying Rare Species}

Rare species for which there was no response option available were the second most common issue. 
Volunteers currently either classify the subject as blank or select a different type of animal from the available options, leading to low consensus.
Despite the benefits of the forced-choice design for reducing non-committal responses and its demonstrated effectiveness for accurate identification \cite{swanson2015snapshot}, when volunteers are unable to indicate that they see a different species than those listed, they may label the image as blank, as demonstrated by the Talk comments presented above.

One design intervention to address this issue would be offering a response option for ``other animal,'' with a contingent text input option for volunteers to name rare species that they observe.
Whether this type of response option would lead to inappropriate use with potentially identifiable animals versus better detection of rare species, would have to be evaluated empirically to establish the impact on classification accuracy.

Another strategy to address this issue is providing volunteers with a more exhaustive list of response options, although existing response options already contain between 53 and 80 species.
The Snapshot Safari science team is wary of presenting an overwhelming number of species to identify, as it could lead to lower consensus levels overall or lower rates of participation due to perceived task difficulty.

Our results showed only small differences in average classification times for projects with larger versus smaller species lists, suggesting that the impacts of more extensive species lists could be relatively minor.
Experimental studies could determine whether the size of the species response option list is a major factor in accuracy or classification efficiency.

A third intervention would be providing options for major species categories that can be expanded to indicate specific species, such as offering Jackal as a top level classification with optional sub-selections of Black-backed jackal or Side-striped jackal.
This strategy has been implemented successfully in platforms such as iNaturalist and eBird, allowing volunteers to identify species to their maximum level of confidence without over-specifying and potentially increasing the frequency of errors \cite{wiggins2016community}.

\subsubsection{Zooming In on Distant Animals}

Another common issue was subjects in which the animal was too far from the camera to easily detect or identify.
Reliably detecting these animals requires using a zoom tool that increases the effort and time required to make a classification, as well as due diligence on the part of volunteers.
This may be a matter of individual differences between volunteers who do or do not take this additional step when a subject does not have animals in the foreground.
Future work could establish whether the use of the zoom or invert image color tools need to be encouraged, as system logging does not currently record these behaviors.

\subsubsection{Clarifying Instructional Content}

The specific problems related to subject quality and target characteristics suggest mixed potential for improvements through updating instructions for volunteers.
Using the project tutorial or other resources (Help buttons, Field Guide for species identification, FAQs) to give volunteers more details on how to handle potential sources of confusion would be among the simplest interventions to implement.
However, the causes of confusion that were most prevalent in our analysis were not a simple matter of clarifying instructions, beyond emphasizing the importance of giving a best guess and zooming in to check if there are distant animals at the horizon or obscured by vegetation.
To keep tutorials streamlined, these instructions would be best implemented in FAQs and discussion forums, where these issues and the best ways to handle them could be described in more detail.
If any of the other potential design interventions identified above were implemented, clear descriptions of when these options should be used would be an important addition to the instructional materials.

\subsection{Summary of Implications}

We summarize our suggested design intervention as follows:  
\begin{itemize}
  \item Develop platform functionality to support dynamic subject retirement rules, as our emergent categories of content complexity indicate potential to retire subjects with fewer classifications when they have either very high or very low agreement.
  \item Provide an option for volunteers to report poor quality images so that these subjects can be retired earlier from the workflow, saving volunteers' time and minimizing potential frustration. 
  \item Explore a response option for ``other animal'' with text input for volunteers to identify and report rare species rather than marking them blank.
  \item Evaluate the strategy of grouping the species options into categories and subcategories, based on taxonomic relationship or visual similarity, to allow volunteers to respond with a more general category when the image content does not support classification to species. 
\end{itemize}
We also suggest future research to evaluate the effectiveness of these design interventions, to develop and test measures of consensus that are more suitable for complex content, and studies that directly evaluate the role of uncertainty in volunteers' classification experiences through interview and survey methodologies.

\section{Conclusion}

To answer our research question \textit{Which features of a subject, such as number or diversity of species, lead to volunteer uncertainty, contribute to low levels of consensus on image classifications, and impact volunteer efficiency?} 
we conducted trace ethnography with mixed methods analysis of six Snapshot Safari projects to understand content characteristics that can lead to uncertainty and low consensus.
Numerous factors can lead volunteers to disagree, including image quality problems and the specifics of interface design and interactions, as well as the complexity of the content itself.

This study contributes a content categorization based on aggregate classifications that characterize image content \textit{complexity} which can contribute to volunteer uncertainty and impact classification efficiency, and which are likely to have parallels in other image classification tasks.
Second, our analysis of how these categories impact volunteers' efficiency in making classifications confirmed their validity.
Third, we developed a schema of image content \textit{characteristics and quality issues} that may impair consensus on classification tasks, and are also likely to occur in other camera trap classification projects.
The results of this study provide empirical evidence to support informed decision-making for prioritizing design interventions and compensatory analyses when using the resulting data.
It also contributes recommendations for future research and system design considerations to support consensus and data quality in image classification tasks.

While our results are specific to the Zooniverse platform and volunteer-supported image classification, similar challenges affect image classification tasks for other platforms and participation models, since issues with image quality and target characteristics, as well as variability in image complexity, are generalizable.
The potential design solutions we identified may be appropriate responses for other platforms as well, though they would similarly benefit from further study to evaluate impacts on data quality and classification efficiency.

Finally, our results indicate a need for further research, particularly experimental studies that assess the impact of alternate interface options.
In the bigger picture, while the loss of a relatively small amount of data due to low consensus may be an acceptable price to pay for crowdsourced classification of camera trap images, the impact on volunteers' participation experiences in citizen science remains a concern that offers numerous opportunities for future research.

\section{Acknowledgments}

This work was supported in part by the Nebraska Research Initiative.
We acknowledge formative work by Doris Uwaezuoke and Shivani Mudhelli, who contributed to earlier stages of this research.
We thank the Snapshot Safari science team and volunteers for sharing their data.
We also thank the anonymous reviewers from the CHI, HCOMP, and CSCW conferences, who provided valuable suggestions for improving this work.

\bibliographystyle{unsrt}  
\bibliography{references}  

\begin{thebibliography}{10}

\bibitem{Wiggins2015}
Andrea Wiggins and Kevin Crowston.
\newblock Surveying the citizen science landscape.
\newblock {\em First Monday}, 20(1), 2015.

\bibitem{rosser2019}
Holly Rosser and Andrea Wiggins.
\newblock Crowds and camera traps: Genres in online citizen science projects.
\newblock In {\em Proceedings of the 52nd Hawaii International Conference on
  System Sciences}, 2019.

\bibitem{trouille2019citizen}
Laura Trouille, Chris~J Lintott, and Lucy~F Fortson.
\newblock Citizen science frontiers: Efficiency, engagement, and serendipitous
  discovery with human--machine systems.
\newblock {\em Proceedings of the National Academy of Sciences},
  116(6):1902--1909, 2019.

\bibitem{rowcliffe2008}
J~Marcus Rowcliffe and Chris Carbone.
\newblock Surveys using camera traps: Are we looking to a brighter future?
\newblock {\em Animal Conservation}, 11(3):185--186, 2008.

\bibitem{mccallum2013}
Jamie McCallum.
\newblock Changing use of camera traps in mammalian field research: Habitats,
  taxa and study types.
\newblock {\em Mammal Review}, 43(3):196--206, 2013.

\bibitem{Yousif_Yuan_Kays_He_2019}
Hayder Yousif, Jianhe Yuan, Roland Kays, and Zhihai He.
\newblock Animal scanner: Software for classifying humans, animals, and empty
  frames in camera trap images.
\newblock {\em Ecology and Evolution}, 9(4):1578–1589, 2019.

\bibitem{Willi_Pitman_Cardoso_Locke_Swanson_Boyer_Veldthuis_Fortson_2019}
Marco Willi, Ross~T. Pitman, Anabelle~W. Cardoso, Christina Locke, Alexandra
  Swanson, Amy Boyer, Marten Veldthuis, and Lucy Fortson.
\newblock Identifying animal species in camera trap images using deep learning
  and citizen science.
\newblock {\em Methods in Ecology and Evolution}, 10(1):80–91, 2019.

\bibitem{Anton_Hartley_Geldenhuis_Wittmer_2018}
Victor Anton, Stephen Hartley, Andre Geldenhuis, and Heiko~U Wittmer.
\newblock Monitoring the mammalian fauna of urban areas using remote cameras
  and citizen science.
\newblock {\em Journal of Urban Ecology}, 4(1), Jan 2018.

\bibitem{ahuja_impact_2019}
Vinod Ahuja, Andrea Wiggins, and Shivani Mudhelli.
\newblock The {Impact} of {Screen} {Size} on {Crowdsourced} {Image}
  {Classification}.
\newblock In {\em Conference {Companion} {Publication} of the 2019 on
  {Computer} {Supported} {Cooperative} {Work} and {Social} {Computing}}, pages
  127--131, Austin TX USA, November 2019. ACM.

\bibitem{beck2018}
Melanie~R Beck, Claudia Scarlata, Lucy~F Fortson, Chris~J Lintott, BD~Simmons,
  Melanie~A Galloway, Kyle~W Willett, Hugh Dickinson, Karen~L Masters, Philip~J
  Marshall, et~al.
\newblock Integrating human and machine intelligence in galaxy morphology
  classification tasks.
\newblock {\em Monthly Notices of the Royal Astronomical Society},
  476(4):5516--5534, 2018.

\bibitem{wright2017}
Darryl~E Wright, Chris~J Lintott, Stephen~J Smartt, Ken~W Smith, Lucy Fortson,
  Laura Trouille, Campbell~R Allen, Melanie Beck, Mark~C Bouslog, Amy Boyer,
  et~al.
\newblock A transient search using combined human and machine classifications.
\newblock {\em Monthly Notices of the Royal Astronomical Society},
  472(2):1315--1323, 2017.

\bibitem{Swanson_Kosmala_Lintott_Packer_2016}
Alexandra Swanson, Margaret Kosmala, Chris Lintott, and Craig Packer.
\newblock A generalized approach for producing, quantifying, and validating
  citizen science data from wildlife images.
\newblock {\em Conservation Biology}, 30(3):520–531, 2016.

\bibitem{Hsing_Bradley_Kent_Hill_Smith_Whittingham_Cokill_Crawley_Stephens_2018}
Pen-Yuan Hsing, Steven Bradley, Vivien~T. Kent, Russell~A. Hill, Graham~C.
  Smith, Mark~J. Whittingham, Jim Cokill, Derek Crawley, and Philip~A.
  Stephens.
\newblock Economical crowdsourcing for camera trap image classification.
\newblock {\em Remote Sensing in Ecology and Conservation}, 4(4):361–374,
  2018.

\bibitem{Law_Gajos_Wiggins_Gray_Williams_2017}
Edith Law, Krzysztof~Z. Gajos, Andrea Wiggins, Mary~L. Gray, and Alex Williams.
\newblock Crowdsourcing as a tool for research: Implications of uncertainty.
\newblock In {\em Proceedings of the 2017 ACM Conference on Computer Supported
  Cooperative Work and Social Computing}, page 1544–1561. ACM, Feb 2017.

\bibitem{anton2018monitoring}
Victor Anton, Stephen Hartley, Andre Geldenhuis, and Heiko~U Wittmer.
\newblock Monitoring the mammalian fauna of urban areas using remote cameras
  and citizen science.
\newblock {\em Journal of Urban Ecology}, 4(1):juy002, 2018.

\bibitem{Tu_Yu_Domeniconi_Wang_Xiao_Guo_2020}
Jinzheng Tu, Guoxian Yu, Carlotta Domeniconi, Jun Wang, Guoqiang Xiao, and
  Maozu Guo.
\newblock Multi-label crowd consensus via joint matrix factorization.
\newblock {\em Knowledge and Information Systems}, 62(4):1341–1369, Apr 2020.

\bibitem{Jung_Lease_2011}
Hyun~Joon Jung and Matthew Lease.
\newblock Improving consensus accuracy via z-score and weighted voting.
\newblock In {\em In Proceedings of the 3rd Human Computation Workshop (HCOMP)
  at AAAI}, page 88–90, 2011.

\bibitem{Sheng_Provost_Ipeirotis_2008}
Victor~S. Sheng, Foster Provost, and Panagiotis~G. Ipeirotis.
\newblock Get another label? improving data quality and data mining using
  multiple, noisy labelers.
\newblock In {\em Proceeding of the 14th ACM SIGKDD international conference on
  Knowledge discovery and data mining - KDD 08}, page 614. ACM Press, 2008.

\bibitem{Sheshadri_Lease}
Aashish Sheshadri and Matthew Lease.
\newblock Square: A benchmark for research on computing crowd consensus.
\newblock page~9.

\bibitem{Williams_Goh_Willis_Ellison_Brusuelas_Davis_Law}
Alex~C Williams, Joslin Goh, Charlie~G Willis, Aaron~M Ellison, James~H
  Brusuelas, Charles~C Davis, and Edith Law.
\newblock Deja vu: Characterizing worker reliability using task consistency.
\newblock page~9.

\bibitem{Zhang_Sheng_Li_Wu_Wu_2017}
Jing Zhang, Victor~S. Sheng, Qianmu Li, Jian Wu, and Xindong Wu.
\newblock Consensus algorithms for biased labeling in crowdsourcing.
\newblock {\em Information Sciences}, 382–383:254–273, Mar 2017.

\bibitem{karger2011iterative}
David~R Karger, Sewoong Oh, and Devavrat Shah.
\newblock Iterative learning for reliable crowdsourcing systems.
\newblock In {\em Advances in neural information processing systems}, pages
  1953--1961, 2011.

\bibitem{de2014crowdsourcing}
A~Garc{\i}a~Seco de~Herrera, Antonio Foncubierta-Rodr{\i}guez, Dimitrios
  Markonis, Roger Schaer, and Henning M{\"u}ller.
\newblock Crowdsourcing for medical image classification.
\newblock In {\em Annual congress SGMI}, volume 2014, 2014.

\bibitem{Schaekermann_Goh_Larson_Law_2018}
Mike Schaekermann, Joslin Goh, Kate Larson, and Edith Law.
\newblock Resolvable vs. irresolvable disagreement: A study on worker
  deliberation in crowd work.
\newblock {\em Proceedings of the ACM on Human-Computer Interaction},
  2(CSCW):1–19, Nov 2018.

\bibitem{Steger_Butt_Hooten_2017}
Cara Steger, Bilal Butt, and Mevin~B. Hooten.
\newblock Safari science: assessing the reliability of citizen science data for
  wildlife surveys.
\newblock {\em Journal of Applied Ecology}, 54(6):2053–2062, 2017.

\bibitem{Clare_Townsend_Anhalt-Depies_Locke_Stenglein_Frett_Martin_Singh_Deelen_Zuckerberg_2019}
John D.~J. Clare, Philip~A. Townsend, Christine Anhalt-Depies, Christina Locke,
  Jennifer~L. Stenglein, Susan Frett, Karl~J. Martin, Aditya Singh, Timothy
  R.~Van Deelen, and Benjamin Zuckerberg.
\newblock Making inference with messy (citizen science) data: when are data
  accurate enough and how can they be improved?
\newblock {\em Ecological Applications}, 29(2):e01849, 2019.

\bibitem{Aceves-Bueno_Adeleye_Feraud_Huang_Tao_Yang_Anderson_2017}
Eréndira Aceves-Bueno, Adeyemi~S. Adeleye, Marina Feraud, Yuxiong Huang,
  Mengya Tao, Yi~Yang, and Sarah~E. Anderson.
\newblock The accuracy of citizen science data: A quantitative review.
\newblock {\em The Bulletin of the Ecological Society of America},
  98(4):278–290, Oct 2017.

\bibitem{Eveleigh_Jennett_Blandford_Brohan_Cox_2014}
Alexandra Eveleigh, Charlene Jennett, Ann Blandford, Philip Brohan, and Anna~L.
  Cox.
\newblock Designing for dabblers and deterring drop-outs in citizen science.
\newblock page 2985–2994. ACM Press, 2014.

\bibitem{Dayan_Gal_Segal_Shani_Cavalier}
Na’ama Dayan, Kobi Gal, Avi Segal, Guy Shani, and Darlene Cavalier.
\newblock Intelligent recommendations for citizen science.
\newblock page~8.

\bibitem{Segal_Gal_Simpson_Victoria_Homsy_Hartswood_Page_Jirotka_2015}
Avi Segal, Ya’akov~(Kobi) Gal, Robert~J. Simpson, Victoria Victoria~Homsy,
  Mark Hartswood, Kevin~R. Page, and Marina Jirotka.
\newblock Improving productivity in citizen science through controlled
  intervention.
\newblock In {\em Proceedings of the 24th International Conference on World
  Wide Web}, page 331–337. ACM, May 2015.

\bibitem{jackson2018}
Corey~Brian Jackson, Kevin Crowston, and Carsten {\O}sterlund.
\newblock Did they login?: Patterns of anonymous contributions in online
  communities.
\newblock {\em Proceedings of the ACM International Conference on
  Computer-supported cooperative work and social computing}, 2(CSCW), 2018.

\bibitem{Jackson_Crowston_Mugar_Osterlund_2016}
Corey~Brian Jackson, Kevin Crowston, Gabriel Mugar, and Carsten \O{}sterlund.
\newblock "guess what! you're the first to see this event": Increasing
  contribution to online production communities.
\newblock In {\em Proceedings of the 19th International Conference on
  Supporting Group Work}, GROUP '16, page 171–179, New York, NY, USA, 2016.
  Association for Computing Machinery.

\bibitem{Mugar_Osterlund_Hassman_Crowston_Jackson_2014}
Gabriel Mugar, Carsten Østerlund, Katie~DeVries Hassman, Kevin Crowston, and
  Corey~Brian Jackson.
\newblock Planet hunters and seafloor explorers: Legitimate peripheral
  participation through practice proxies in online citizen science.
\newblock In {\em Proceedings of the 17th ACM Conference on Computer Supported
  Cooperative Work 38; Social Computing}, CSCW ’14, page 109–119. ACM,
  2014.

\bibitem{bowyer2015panoptes}
Alex Bowyer, Chris Lintott, Greg Hines, Campbell Allan, and Ed~Paget.
\newblock Panoptes, a project building tool for citizen science.
\newblock In {\em Proceedings of the AAAI Conference on Human Computation and
  Crowdsourcing (HCOMP’15). AAAI, San Diego, CA, USA}, pages 1--2, 2015.

\bibitem{swanson2015snapshot}
Alexandra Swanson, Margaret Kosmala, Chris Lintott, Robert Simpson, Arfon
  Smith, and Craig Packer.
\newblock Snapshot serengeti, high-frequency annotated camera trap images of 40
  mammalian species in an african savanna.
\newblock {\em Scientific data}, 2:150026, 2015.

\bibitem{hines2015aggregating}
Greg Hines, Alexandra Swanson, Margaret Kosmala, and Chris Lintott.
\newblock Aggregating user input in ecology citizen science projects.
\newblock In {\em Twenty-Seventh IAAI Conference}, 2015.

\bibitem{jackson2016way}
Corey Jackson, Carsten {\O}sterlund, Veronica Maidel, Kevin Crowston, and
  Gabriel Mugar.
\newblock Which way did they go? newcomer movement through the zooniverse.
\newblock In {\em Proceedings of the 19th ACM conference on computer-supported
  cooperative work \& social computing}, pages 624--635, 2016.

\bibitem{neuendorf}
Kimberly~A Neuendorf.
\newblock {\em The content analysis guidebook}.
\newblock Sage, 2016.

\bibitem{wiggins2016community}
Andrea Wiggins and Yurong He.
\newblock Community-based data validation practices in citizen science.
\newblock In {\em Proceedings of the 19th ACM Conference on computer-supported
  cooperative work \& social computing}, pages 1548--1559, 2016.

\end{thebibliography}

\end{document}